\documentclass[]{aastex701}

\usepackage{xspace}

\newcommand{\pnp}[1]{\textcolor{black}{#1}}
\newcommand{\rr}[1]{\textcolor{black}{#1}}

\newcommand{\rprocess}{$r$-process\xspace}
\newcommand{\unit}[1]{\ensuremath{\mathrm{\,#1}}\xspace}
\newcommand{\Msun}{\unit{M}_\odot}

\begin{document}

\title{Implications of low neutron star merger rates for gamma-ray bursts, r-process production and Galactic double neutron stars}

\author[0000-0002-1980-5293]{Maya Fishbach}
\email{maya.fishbach@utoronto.ca}
\affiliation{Canadian Institute for Theoretical Astrophysics, 60 St George St, University of Toronto, Toronto, ON M5S 3H8, Canada}
\affiliation{David A. Dunlap Department of Astronomy and Astrophysics, University of Toronto, 50 St George St, Toronto ON M5S 3H4, Canada}
\affiliation{Department of Physics, 60 St George St, University of Toronto, Toronto, ON M5S 3H8, Canada}

\author[0000-0002-4863-8842]{Alexander~P.~Ji}
\email{alexji@uchicago.edu}
\affiliation{Department of Astronomy \& Astrophysics, University of Chicago, 5640 S Ellis Avenue, Chicago, IL 60637, USA}
\affiliation{Kavli Institute for Cosmological Physics, University of Chicago, Chicago, IL 60637, USA}
\affiliation{NSF-Simons AI Institute for the Sky (SkAI), 172 E. Chestnut St., Chicago, IL 60611, USA}

\author[0000-0002-7374-935X]{Wen-fai Fong} 
\email{wfong@northwestern.edu}
\affiliation{Center for Interdisciplinary Exploration and Research in Astrophysics (CIERA) and Department of Physics and Astronomy, Northwestern University, Evanston, IL 60208, USA}

\author[0009-0009-7362-4758]{Tom Y. Wu} 
\email{tomwu@unc.edu}
\affiliation{Department of Physics and Astronomy, University of North Carolina at Chapel Hill, 120 E. Cameron Ave, Chapel Hill, NC, 27599,
USA}
\affiliation{David A. Dunlap Department of Astronomy and Astrophysics, University of Toronto, 50 St George St, Toronto ON M5S 3H4, Canada}

\author[0000-0002-9267-6213]{Jillian C. Rastinejad} 
\email{jcrastin@umd.edu}
\altaffiliation{NHFP Einstein Fellow}
\affiliation{Department of Astronomy, University of Maryland, College Park, MD~20742, USA}

\author[0000-0002-4103-0666]{Aditya Vijaykumar}
\email{aditya@utoronto.ca}
\affiliation{Canadian Institute for Theoretical Astrophysics, 60 St George St, University of Toronto, Toronto, ON M5S 3H8, Canada}

\author[0000-0001-5403-3762]{Hsin-Yu Chen}
\email{hsinyu@austin.utexas.edu}
\affiliation{Department of Physics, The University of Texas at Austin, 2515 Speedway, Austin, TX 78712, USA}

\begin{abstract}
The first multimessenger discovery of a binary neutron star (BNS) merger, GW170817, proved that such mergers can source short gamma-ray bursts (SGRBs) and produce \rprocess elements. The initial merger rate from this single event was found to be broadly consistent with the SGRB rate, the Milky Way (MW) \rprocess mass, and the Galactic population of double neutron star (DNS) systems that will merge in a Hubble time. However, only one additional BNS merger has been detected since, and the BNS merger rate has been consistently revised downwards with recent gravitational wave (GW) catalog updates. Analyzing GWTC-4, we find a total BNS merger rate of $28$--$300\,\mathrm{Gpc}^{-3}\,\mathrm{yr}^{-1}$ consisting of $53^{+176}_{-49}\,\mathrm{Gpc}^{-3}\mathrm{yr}^{-1}$ in GW170817-like $\sim(1.3,1.3)\,M_\odot$ BNSs (90\%~credibility). We revisit the consistency of the BNS merger rate with SGRBs, \rprocess and Galactic DNSs. In all cases, there is an emerging tension with the BNS (and EM-bright neutron star--black hole, NSBH) merger rate. Comparing to a BNS merger rate of $100\,\mathrm{Gpc}^{-3}\mathrm{yr}^{-1}$, the cosmological SGRB rate is a factor of 3.6--18 higher \rr{(despite kilonova followup of SGRBs implying a significant fraction of SGRBs are of BNS origin)}, the \rprocess rate is a factor of 0.9--4.1 higher \rr{(even though we consider only \rprocess elements above the second peak)}, and the rate inferred from Galactic DNSs is a factor of 2.3--5.1 higher than the BNS rate. We discuss how various uncertainties in the inferred rates either alleviate or exacerbate this tension, which point to the various physical processes that can be constrained by such rate comparisons.
\end{abstract}

\section{Introduction}

The LIGO-Virgo-KAGRA (LVK) gravitational-wave (GW) observatory network~\citep{2015CQGra..32g4001L,2015CQGra..32b4001A,2021PTEP.2021eA101A} has recently concluded its fourth observing run, yielding hundreds of new 
GW observations.\footnote{ \url{https://gracedb.ligo.org/superevents/public/O4/}.} 
The latest GW transient catalog, GWTC-4, covers the first 8 months of the 2.5-year observing run, and includes around 100 new binary black hole (BBH) candidates and a few neutron star--black hole (NSBH) binaries, bringing the total number of published GW events to $\approx200$~\citep{LIGOScientific:2025slb}. 
However, no significant binary neutron star (BNS) events have been reported in the first part of the fourth observing run.\footnote{A subthreshold BNS candidate was reported in \citet{Niu:2025nha} but because its astrophysical origin is uncertain due to its subthreshold nature, we do not include it in this study.} 
Furthermore, none of the reported NSBH events in GWTC-4 had corresponding electromagnetic emission (EM) -- neither observed nor theorized, because the masses and spins of the detected NSBH systems are unlikely to result in neutron star (NS) disruption outside the black hole (BH) horizon~\citep{LIGOScientific:2024elc,LIGOScientific:2025slb}. 

The only high-significance BNS events detected with GWs remain GW170817~\citep{LIGO2017} and GW190425~\citep{LIGO2020}. The first BNS event GW170817 is also the first multimessenger GW detection, with counterpart emission observed across the EM spectrum, enabling tremendous discoveries across nuclear physics, gravitational physics, high-energy astrophysics and cosmology~\citep{LIGOScientific:2017ync}. 
GW170817 definitively proved the long-standing hypotheses that some gamma-ray bursts (GRBs) originate in BNS mergers~\citep{Berger:2013jza,LIGOScientific:2017zic,2017ApJ...848L..14G} and BNS mergers are responsible for the production of some \rprocess elements, powering transients known as kilonovae~\citep{Metzger:2019zeh,DES:2017kbs,2017Natur.551...80K}.
The second BNS event GW190425 likely also had some corresponding EM emission (potentially powering a GRB and/or kilonova), but it was never observed, potentially owing to its greater distance and poor GW sky localization~\citep{Keinan:2024gai,2025ApJ...988..169C}. 
NSBH events, a handful of which have been observed by the LVK, can also power GRBs and kilonovae, but this requires a relatively low-mass and/or spinning BH component with a small innermost stable circular orbits (ISCO) so that the NS can tidally disrupt and leave behind a remnant before falling into the BH, with details depending on uncertain the NS equation of state~\citep{Foucart:2018rjc}. 
None of the NSBH events observed so far are likely to meet these criteria, and it is therefore improbable that they had EM counterparts~\citep{Biscoveanu:2022iue,Biscoveanu_2023}. The most promising candidate for an EM-bright NSBH is GW230529 because of its low-mass BH component, but the probability it experienced NS tidal disruption is less than 10\%~\citep{LIGOScientific:2024elc}, \pnp{and indeed no EM counterpart was identified~\citep{2024ApJ...970L..20R,2025PhRvD.112h3002P}.} 

Despite the shortage of NS mergers (NSM) with detected EM and GW emission, we can still study the association between GW and EM phenomena on a population level.
\pnp{Merger-origin GRBs inform the population properties of NSMs independently of directly observing GW counterparts~\citep{Chen:2012qh,
2020ApJ...893...38B,2020ApJ...895..108F,2022PhRvD.105h3004S,2022A&A...666A.174S,Chen:2025otp,Kunnumkai2026}.}
Likewise, studying \rprocess abundances constrains the NSM populations that produce these elements~\citep{Zevin:2019obe,Holmbeck:2023xqy,Frebel:2023hsk}.
Another probe of the BNS merger population is the population of double neutron star (DNS) systems in our galaxy, of which there are a couple dozen observed systems~\citep{Ozel2012,2018MNRAS.478.1377A,Pol2019,2020RNAAS...4...65F,2025arXiv251212130A}.

\begin{figure}
    \centering
    \includegraphics[width=\linewidth]{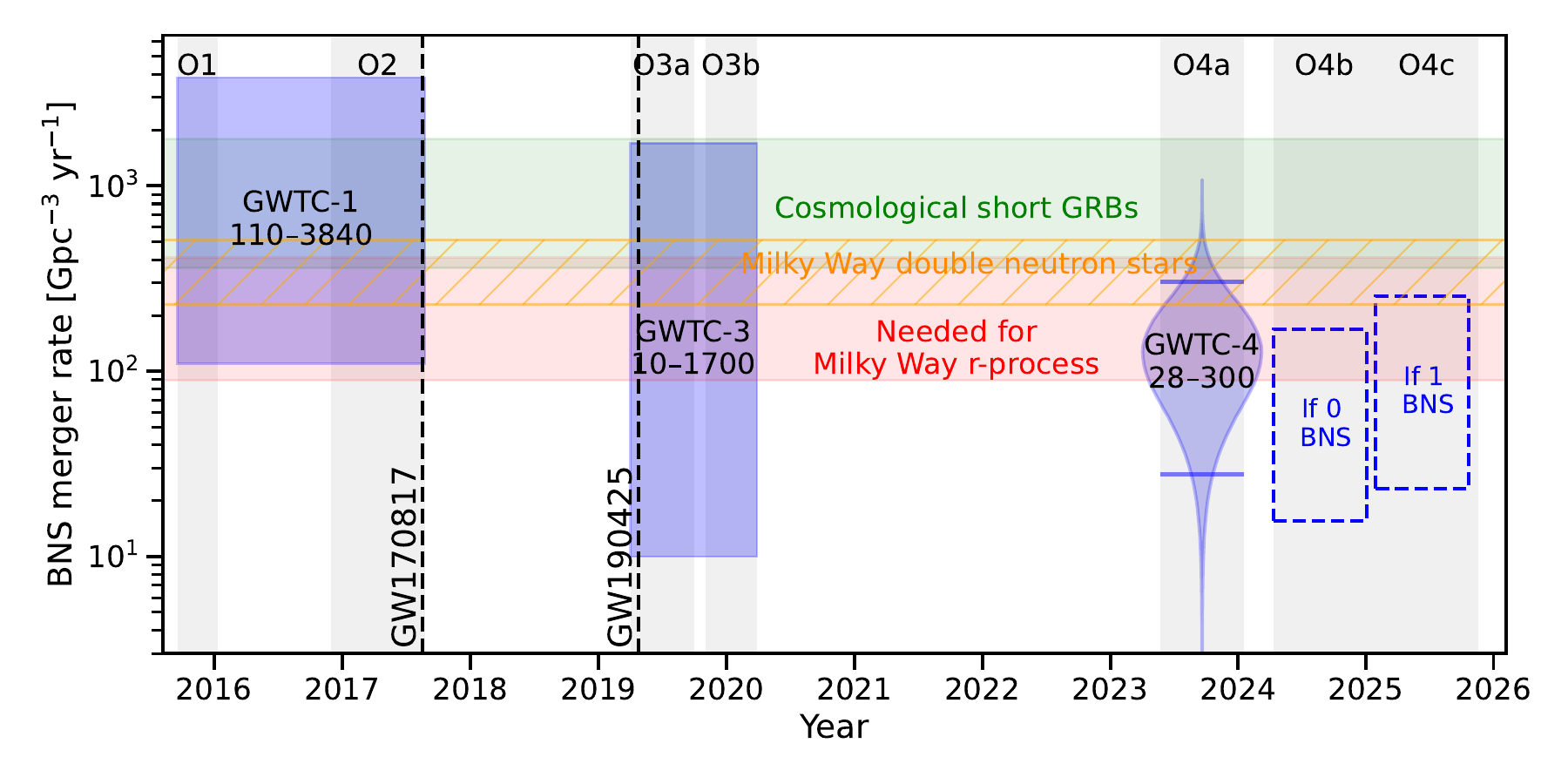}
    \caption{Inferred BNS merger rate as a function of time, compared to the rates of short GRBs, MW DNS systems, and the rate required to explain the MW \rprocess mass. The GWTC-1 and GWTC-3 inferred rates show the union of 90\% error bars over different posteriors (that assume different mass distributions), so we represent them as shaded blue rectangles. The GWTC-4 posterior inferred in this work is shown as a violin. The dashed, unfilled blue rectangles show the approximate 90\% credible intervals on the BNS rate at the end of O4 if there are 0 new BNS or 1 new BNS detected in O4 once the data is fully analyzed.}
    \label{fig:rate-comparisons}
\end{figure}

Following GW170817, the BNS merger rate was inferred to be $1540^{+3200}_{-1220}\,\mathrm{Gpc}^{-3}\,\mathrm{yr}^{-1}$ at 90\% credibility~\citep{LIGO2017}. The inferred merger rate depends on the BNS mass distribution, and updated analyses considering different mass distributions found a merger rate of $110$--$3840\,\mathrm{Gpc}^{-3}\,\mathrm{yr}^{-1}$ following the second LVK observing run~\citep{LIGOScientific:2018mvr} \pnp{and $10$--$1700\,\mathrm{Gpc}^{-3}\,\mathrm{yr}^{-1}$ following the third LVK observing run~\citep{KAGRA:2021duu}.} 
Following GWTC-4 and the decreased detection rate of BNS mergers despite the increase in detector sensitivity, the inferred astrophysical merger rate has been revised downwards to $7.6$--$250\,\mathrm{Gpc}^{-3}\,\mathrm{yr}^{-1}$, again marginalizing over different mass distributions~\citep{LIGOScientific:2025pvj}.
In this work, we provide a conservative estimate for the BNS merger rate under slightly different assumptions (erring on the side of a higher rate), finding $28$--$300\,\mathrm{Gpc}^{-3}\,\mathrm{yr}^{-1}$ for GWTC-4. The inferred BNS merger rate as a function of time is summarized in Fig.~\ref{fig:rate-comparisons}.

Fig.~\ref{fig:rate-comparisons} also includes simplified projections for the BNS rate inferred at the end of O4, assuming no new BNS are observed or one new BNS  is observed. These projections are estimated assuming that the total number of event candidates (as released in public alerts) is proportional to the total surveyed spacetime volume. The end of O4 had 391 cumulative public alerts and/or published events compared to 218 at the end of O4a. This implies that the surveyed spacetime volume at the end of O4 is approximately $391/218$, or 1.8, times larger than the surveyed spacetime volume at the end of O4a. If no new BNS mergers are detected, the rate at the end of O4 would be roughly 1.8 times smaller than the O4a rate. If one new BNS merger is detected, the number of BNS events would increase by a factor of 1.5 (from 2 to 3 detections), so the end-of-O4 rate would be roughly $1.5/1.8$, or 80\%, of the O4a rate.

Observations of GRBs, \rprocess abundance measurements, and the Galactic DNS population each provide a measurement of the BNS merger rate that is independent of GW observations. 
\pnp{Prior to the latest LVK catalog GWTC-4}, the GW-inferred BNS merger rates of a few hundred $\mathrm{Gpc}^{-3}\,\mathrm{yr}^{-1}$ were consistent with these independent estimates~\citep{MandelBroekgaarden2022}. In this work, we revisit these comparisons in light of the lower BNS merger rate, identifying possible tensions between different rate estimates and their implications.  
The remainder of this paper is structured as follows. 
In \S~\ref{sec:GWrates}, we infer the merger rates of BNS and ``EM-bright" NSBH merger rates from the most recent GW catalog GWTC-4. We simplify the calculation by reporting merger rates in specific mass bins, where each mass bin corresponds to at most one GW event. 
Our merger rate is therefore relatively robust to any mass distribution modeling systematics and extrapolation effects, although we discuss the sensitivity of the inferred rate to NS mass and spin distribution assumptions. 
Summing over all masses, we infer the total BNS rate to be $110^{+192}_{-82}\,\mathrm{Gpc}^{-3}\,\mathrm{yr}^{-1}$.
In \S\ref{sec:GRB}, we compile estimates of the merger-origin GRB rate, which tend to be higher than the inferred BNS merger rate, and discuss implications for GRB progenitors, beaming angles and cosmological evolution.
In \S\ref{sec:rprocess}, we derive the \rprocess mass in the Milky Way (MW) and the implied BNS merger rate needed to produce this mass, comparing it to the GW rates and discussing implications for \rprocess production. 
In \S\ref{sec:MW-DNS}, we compare our inferred BNS merger rate as a function of mass to the Galactic double neutron star (DNS) population and discuss implications for pulsar beaming and radio survey selection effects. 
We discuss remaining uncertainties and implications for future observations in \S\ref{sec:discussion} and summarize in \S\ref{sec:conclusion}.

\section{Neutron star merger rates inferred from GWTC-4}
\label{sec:GWrates}
In a companion paper to the latest GW catalog, GWTC-4, \citet{LIGOScientific:2025pvj} simultaneously fit for the compact object mass distribution and the merger rates of BNS, NSBH and BBH. Here, we report the rates of BNS and low-mass NSBH mergers in specific mass bins, recovering total rates consistent with \citet{LIGOScientific:2025pvj}. We also discuss the sensitivity of these rate estimates to assumptions about the NS mass and spin distributions. 

\subsection{Merger rates in mass bins}

\begin{figure}
    \centering
    \includegraphics[width=0.5\linewidth]{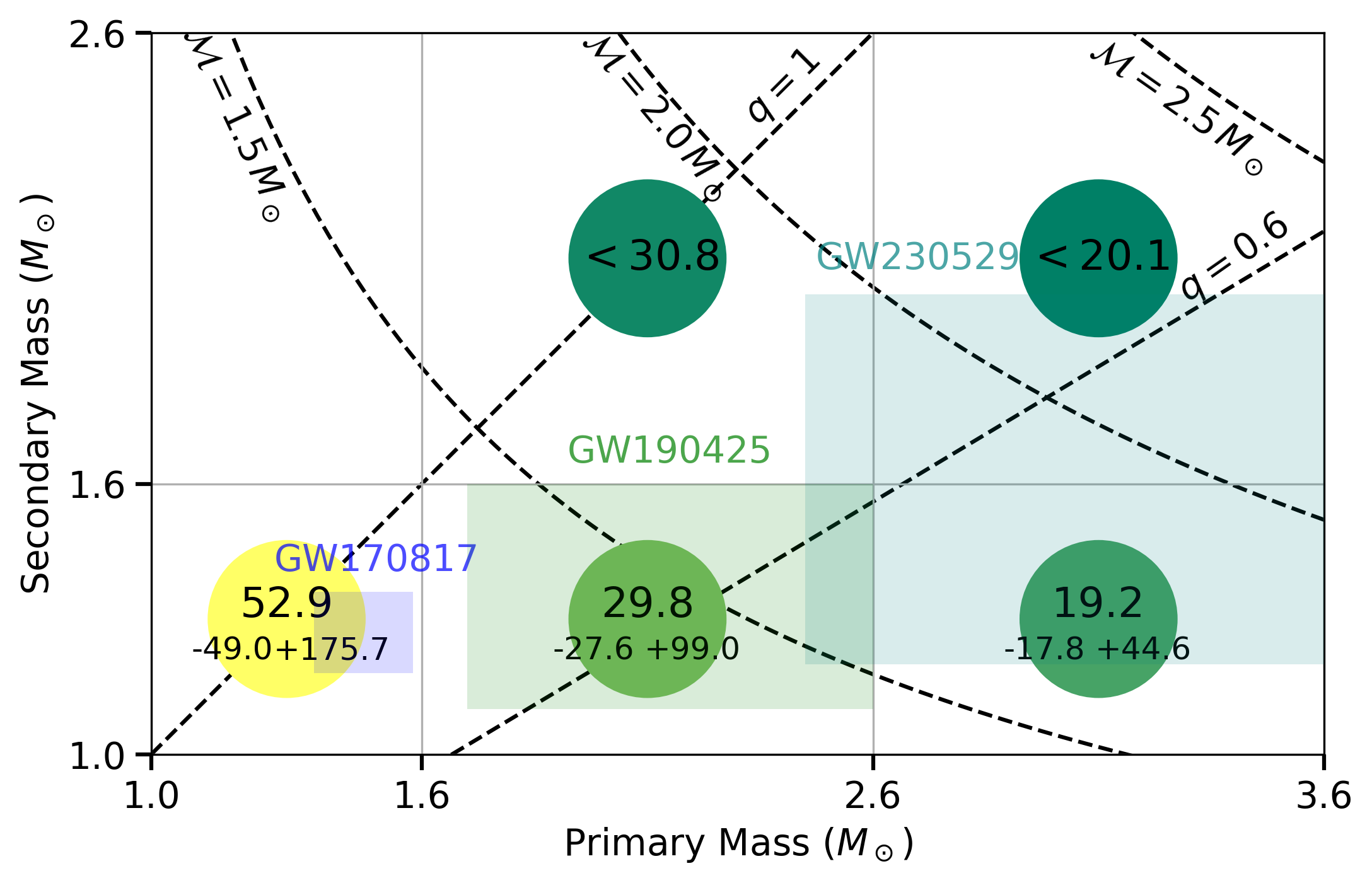}
    \caption{BNS and low-mass NSBH rates, in units of $\mathrm{Gpc}^{-3}\mathrm{yr}^{-1}$, inferred from the latest GW catalog, GWTC-4, as a function of their component masses, taking mass bins with edges of 1, 1.6, 2.6 and 3.6 $M_\odot$. For mass ranges at which no mergers have been detected, we report 90\% upper limits. For mass ranges that have at least one detection, we report the median and 90\% credibility interval. We conservatively count GW230529 in the bin centered at $1.3+3\,M_\odot$, although its BH component may be as high as $4.5\,M_\odot$. We assume the mass distribution is uniform in each bin, with a uniform spin distribution up to 0.4 for masses below $2.6\,M_\odot$, and a uniform spin distribution up to 1 for masses above $2.6\,M_\odot$. The joint BNS merger rate (median and 90\% credible interval) is $110_{-82}^{+192}\mathrm{Gpc}^{-3}\mathrm{yr}^{-1}$. The joint BNS/ low-mass NSBH merger rate is $147_{-96}^{+198}\mathrm{Gpc}^{-3}\mathrm{yr}^{-1}$.}
    \label{fig:bns-rates}
\end{figure}

GWTC-4 contains two significant\footnote{For consistency with~\citet{LIGOScientific:2025pvj}, we classify a GW detection as significant if (a) its FAR is lower than 0.25 yr$^{-1}$ in at least one search pipeline; or (b) the probability of astrophysical origin $p_{\text{astro}} > 0.9$ in at least one search pipeline; or (c) the network signal-to-noise ratio SNR $> 10$. We classify NS components as those with mass below $2.3\,M_\odot$ at $>90\%$ credibility under the default parameter estimation posteriors reported in GWTC-4~\citep{LIGOScientific:2025slb}. } BNS mergers, GW170817~\citep{LIGO2017} and GW190425~\citep{LIGO2020}, detected in the LVK's second and third observing runs, respectively. In terms of its masses, GW170817 is consistent with the Galactic merging BNS population (component masses $\approx1.36\,M_\odot$) while GW190425 is more massive (containing one component mass $\gtrsim1.7\,M_\odot$).
Our results exclude the subthreshold BNS candidate GW231109\_235456 recently identified in the first part of the fourth observing run by~\citet{Niu:2025nha} because it does not meet our significance threshold and is therefore incompatible with our sensitivity estimation.
The inferred BNS merger rate is degenerate with the BNS mass distribution, which is difficult to constrain with only two GW events. 
Instead, we infer the BNS merger rate in specific mass bins, equivalent to assuming a piecewise-constant, binned histogram BNS mass distribution. 
Splitting the BNS mass distribution into three bins centered at $1.3\,M_\odot+1.3\,M_\odot$, $1.3\,M_\odot+2.1\,M_\odot$, and $2.1\,M_\odot+2.1\,M_\odot$, we can infer the merger rate in each bin. 
The low mass bin contains GW170817, the second bin contains GW190425, and the third bin contains no events, allowing us to place an upper limit on its corresponding merger rate.
We assume that BNS component masses are uniformly distributed in each bin with bin edges 1--1.6 $M_\odot$ and 1.6--2.6 $M_\odot$.
Within each bin, we assume component masses are randomly paired into binaries.
\rr{This is a similar, but simpler, approach to the ``Binned Gaussian Process" population model~\citep{2023ApJ...957...37R} in \citet{LIGOScientific:2025pvj}; we allow the merger rate in each mass bin to vary independently, instead of enforcing a Gaussian process prior that imposes a smooth covariance across bins.}

Taking into account the scaling of the detection sensitivity with mass -- the surveyed spacetime volume scales roughly as $\mathcal{M}^{5/2}$ where $\mathcal{M}$ is the chirp mass -- the inferred merger rate in each of the three bins is approximately equivalent to the rate of a $\delta$-function mass distribution at $1.30\,M_\odot+1.30\,M_\odot$, $1.65\,M_\odot+1.65\,M_\odot$ and $2.10\,M_\odot+2.10\,M_\odot$. 
We assume BNS component spins follow a uniform spin distribution between dimensionless spins of 0 and 0.4. 
\pnp{We evaluate merger rates following \citet{2025PhRvD.112j2001E} using the cumulative LVK search sensitivity estimates~\citep{LIGO_Scientific_Collaboration2025-kp}.}

We also consider mass ranges corresponding to EM-bright NSBH mergers, which must involve a low-mass BH in order to disrupt the NS outside the BH horizon and potentially produce EM emission.
We take BH component mass bins between $2.6$ and $3.6\,M_\odot$ with uniform BH spins between 0 and 1 for these NSBH, and also assume random component mass pairing within each bins.
The only NSBH event that may fall into one of these bins is GW230529. Although its primary mass may be as high as $4.5\,M_\odot$ (95\% upper limit assuming default parameter estimation priors), if we include it in the 1--$1.6\,M_\odot$, 2.6--$3.6\,M_\odot$ bin, we can get a conservative upper limit on the rate of EM-bright NSBH mergers.
This is a conservative limit because taking into account its masses, spins and realistic NS equation of state, GW230529 has a $<10\%$ probability of powering EM emission.
We take a flat-in-log prior ($p(R) \propto 1/R)$ on the merger rate in each bin, {except for bins with zero detections, for which we take a Jeffreys prior ($p(R) \propto 1/\sqrt{R}$), yielding a conservative upper limit}.

The inferred BNS and low-mass NSBH rates in each mass bin are shown in Fig.~\ref{fig:bns-rates} as either 90\% symmetric credible intervals (in bins containing a detection) or 90\% upper limits (in bins containing zero detections).
The inferred merger rate of BNS involving low, GW170817-like masses is $53^{+176}_{-49}\,\mathrm{Gpc}^{-3}\mathrm{yr}^{-1}$ at 90\% credibility. For slightly higher primary masses (GW190425-like), the inferred merger rate is $30^{+99}_{-28}\,\mathrm{Gpc}^{-3}\mathrm{yr}^{-1}$.
 For BNS containing two high-mass (between $1.6$--$2.6\,M_\odot$) components, we can place a 90\% upper limit of $<30\,\mathrm{Gpc}^{-3}\mathrm{yr}^{-1}$.
The total BNS merger rate is $110_{-82}^{+192}\mathrm{Gpc}^{-3}\mathrm{yr}^{-1}$; this posterior probability density is plotted in Fig.~\ref{fig:rate-comparisons}. Including low-mass NSBHs, the NSM rate is $147_{-96}^{+198}\,\mathrm{Gpc}^{-3}\mathrm{yr}^{-1}$.
This is a conservatively high estimate of the EM-bright NS merger rate because it counts GW230529, which has $<10\%$ probability of being EM-bright, and places a conservative Jeffreys prior on bins with zero detections.
A more accurate rate estimate of the EM-bright NSBH rate should account for the NS equation of state and the population-informed BH mass and spin to calculate the remnant mass and probability of powering EM emission \pnp{(see, e.g., \citealt{Biscoveanu_2023}, who found an upper limit on the EM-bright NSBH rate of $<20\,\mathrm{Gpc}^{-3}\,\mathrm{yr}^{-1}$ using GWTC-3, and \citealt{Kunnumkai2026}, who found consistent results with GWTC-4).}
These rates represent local $z = 0$ merger rates, as current GW detectors are only sensitive to BNS mergers out to $z < 0.1$.
Our inferred BNS merger rate ($28-300\,\mathrm{Gpc}^{-3}\mathrm{yr}^{-1}$ at 90\% credibility) is consistent with the latest published BNS rate from the LVK, which, marginalizing over uncertainties in the mass distribution, found it to be in the range $7.6$--$250\,\mathrm{Gpc}^{-3}\mathrm{yr}^{-1}$~\citep{LIGOScientific:2025pvj}.
\rr{The flexibility of our model (which allows the rate to vary independently in each mass bin) and our prior choices, particularly for bins with zero detections, results in a conservatively high rate estimate with conservatively large error bars.}

\subsection{Sensitivity to mass and spin distributions}

\label{sec:GWrates-systematics}

\begin{figure}
    \centering
    \includegraphics[width=0.5\linewidth]{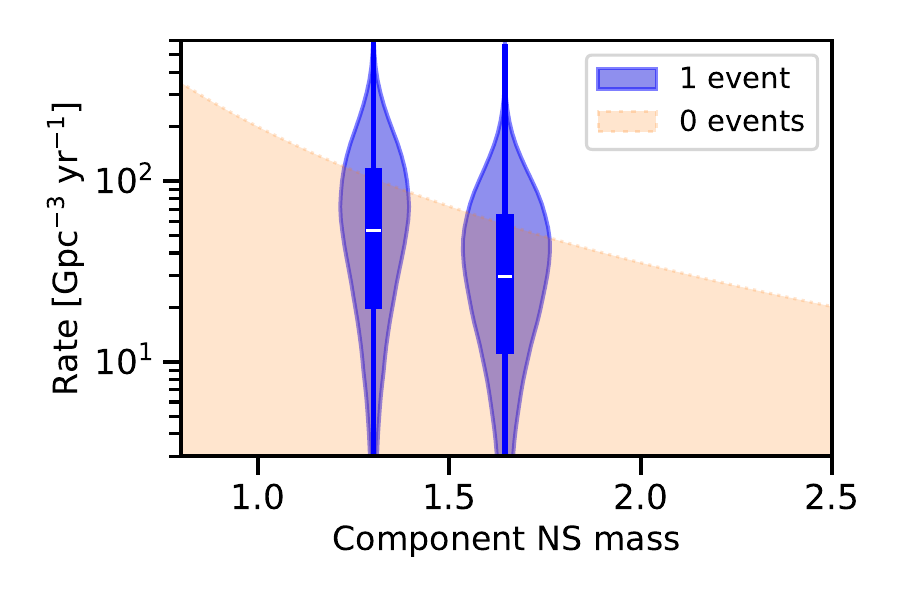}
    \caption{Inferred merger rates for $\delta$-function mass distributions of equal-mass binaries. 
    The posteriors on the merger rates of GW170817-like and GW190425-like mergers are shown as violins (matching their chirp mass to the corresponding component mass of an equal-mass binary). At masses for which there are no detections, we show the 90\% upper limit on the inferred rate under a Jeffreys prior. If the BNS mass distribution skewed to significantly lower masses than we consider (component masses $\lesssim1\,M_\odot$), the merger rate can be at most a factor of $\sim2-3$ higher than our reported merger rate.}
    \label{fig:BNS-rate-v-mass}
\end{figure}

We investigate how the the merger rates inferred in the previous subsection may vary depending on the assumed NS mass and spin distributions.
We assumed a piecewise-flat component mass distribution in three BNS mass bins ranging from 1 to 2.6 $M_\odot$. This gives equivalent rates to $\delta$-function mass distributions of equal-mass binaries centered at $1.3$, $1.65$ and $2.1\,M_\odot$, with one detection in the first two bins and zero detections in the third bin.
We show the inferred merger rate corresponding to various $\delta$-function, equal-mass BNS mass distributions in Fig.~\ref{fig:BNS-rate-v-mass}. The inferred merger rates of GW170817-like and GW190425-like binaries (based on their chirp masses) are shown as violins. 
In orange, we show the 90\% upper limit on the merger rate for masses at which there are no detections, inferred under a Jeffreys prior. 

The sensitive time-volume $VT$ in which we can detect BNS mergers \rr{scales with the chirp mass $\mathcal{M}$ roughly as
\begin{equation}
VT \propto \mathcal{M}^{5/2},
\end{equation}
while the expected number of observed events is $\mu = \mathcal{R} \times VT$, where $\mathcal{R}$ is the merger rate. In other words,
\begin{equation}
\mathcal{R} \propto \mu \mathcal{M}^{-5/2}.
\end{equation}
Meanwhile, the under a Jeffreys prior, the 90\% upper limit on the expected number of events $\mu$ given zero observations is $\mu_{90} = 1.35$. 
Thus, we can place a 90\% upper limit on the merger rate $\mathcal{R}_{90}$ of a hypothetical BNS population of chirp mass $\mathcal{M}$ (in solar masses),
\begin{equation}
     \mathcal{R}_{90} = 1.35\, \mathcal{R}_{\rm GW170817} \left(\frac{\mathcal{M}}{1.3}\right)^{-5/2}
\end{equation}
}
As Fig.~\ref{fig:BNS-rate-v-mass} shows, the contribution of undetected higher-mass BNS to the merger rate is relatively negligible, because their sensitive volume is much larger, while under the extreme assumption that the BNS mass distribution skews to much lower masses than assumed here ($1\,M_\odot$ or below), the BNS merger rate can be at most 2--3 times higher than we reported in the previous subsection.


Our inferred merger rates also assumed that NS spins follow a uniform component spin distribution up to maximum spins of 0.4.
However, the LVK searches are less sensitive to BNS with spins $>0.05$ due to a lack of high-spin templates used for matched filtering~\citep{LIGOScientific:2025yae}. \rr{Using the LVK sensitivity estimates to calculate the sensitive volume of BNS populations with different spin distributions~\citep{LIGO_Scientific_Collaboration2025-kp,2025PhRvD.112j2001E}}, we find that if BNS mergers have spins restricted $<0.05$, the merger rate would be lower by a factor of $\sim1.9$ compared to what we report here, because of the higher detection probability of low-spin BNS mergers. 
Under the extreme assumption that BNS mergers were all restricted to higher spins ($>0.4$), the inferred merger rate would not change by more than a couple percent compared to our reported rate. 
\rr{Hence, our adopted spin distribution yields a conservative (high) estimate of the BNS merger rate. Assuming BNS spins are always below 0.05 would exacerbate the rate tensions that we discuss in the follow sections by a factor of $\sim2$.}

\section{Gamma-ray Bursts}
\label{sec:GRB}

\begin{figure}
    \centering
    \includegraphics[width=0.7\linewidth]{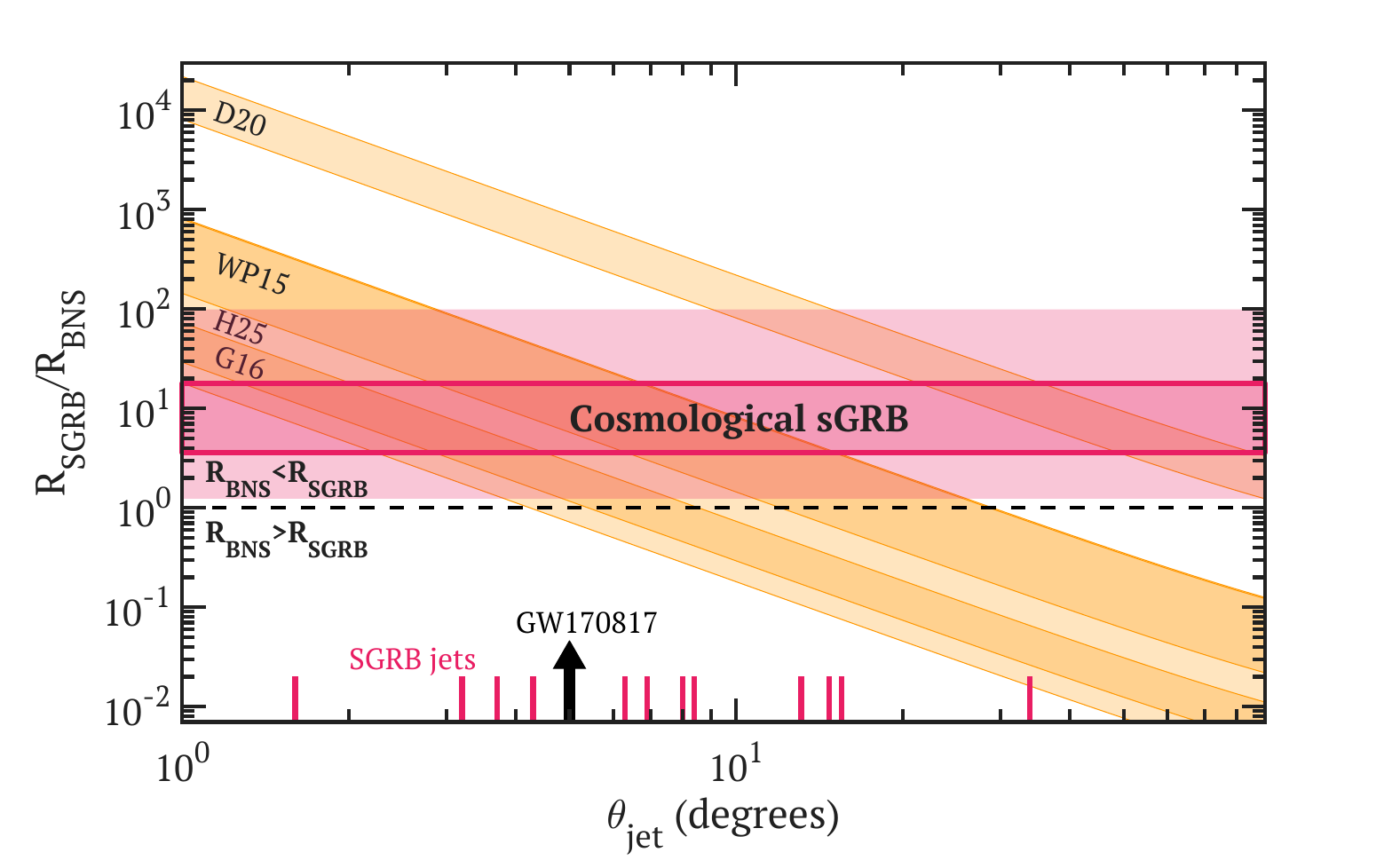}
    \caption{Ratio of volumetric rates $\mathcal{R_{\rm SGRB}}/\mathcal{R_{\rm BNS}}$ versus jet opening angle $\theta_\mathrm{jet}$ in degrees, assuming a central value $\mathcal{R_{\rm BNS}}=100$~Gpc$^{-3}$~yr$^{-1}$. The orange bands represent local rates ($<200$~Mpc) inferred from SGRBs \citep{Wanderman2015,Ghirlanda2016,Dichiara2020,Howell2025} in which the width of the band is set by the uncertainties in both the BNS and SGRB rates. Also shown are the cosmological SGRB rates inferred from SGRB jet opening angles assuming a local rate of 10~Gpc$^{-3}$yr$^{-1}$ \citep{RoucoEscorial2023}. We show two ranges: (i) assuming a fixed central value of $\mathcal{R_{\rm BNS}}=100$~Gpc$^{-3}$~yr$^{-1}$ (central red band) resulting in a cosmological SGRB rate of $\sim 3.6-18$~times larger than the latest BNS rate, and (ii) incorporating the uncertainties of both the BNS and SGRB rates representing a very conservative and wide range, resulting in a ratio of $\sim 1.3-100$. It is also clear that wider jets of $\gtrsim 10^{\circ}$ for most SGRBs would be required to reduce any rate tension, which represents only a minority of SGRB jet opening angles (red lines).
    }
    \label{fig:rates_ratio}
\end{figure}

The coincident detection between GW170817 and a short-duration gamma-ray burst (SGRB) GRB\,170817A demonstrated that at least some BNS mergers produce successful relativistic outflows in the form of SGRBs. With updated BNS merger rates ($\mathcal{R_{\rm BNS}}$; Section~\ref{sec:GWrates}), it is both instructive and timely to compare the BNS rates with those inferred from SGRBs ($\mathcal{R_{\rm SGRB}}$) to understand the fraction of BNS mergers that produce observable SGRBs.  If most or all BNS mergers produce SGRBs, then we expect $\mathcal{R_{\rm BNS}}\sim\mathcal{R_{\rm SGRB}}$, a scenario which is supported by past studies on rates (see \citealt{MandelBroekgaarden2022} for a review). However, the large uncertainties in $\mathcal{R_{\rm SGRB}}$ \citep{Coward2012,Fong2015,Ghirlanda2016,Zhang2018,DellaValle2018}, coupled with the initially higher inferred BNS rates from earlier GW observing runs \citep{LIGO2017,LIGOScientific:2018mvr} have also allowed for the scenario that the BNS rate exceeds the SGRB rate ($\mathcal{R_{\rm BNS}}>\mathcal{R_{\rm SGRB}}$). Indeed, this would imply that some SGRBs could be ``choked'' and/or never break-out of the BNS merger ejecta, for which there is some theoretical support (i.e., \citealt{Murguia-Berthier2014,Nagakura2014,Pavan2025}). On the other hand, if the SGRB rate is larger than the BNS merger rate ($\mathcal{R_{\rm BNS}}<\mathcal{R_{\rm SGRB}}$), this would motivate a revision to the SGRB rate {\it or} a separate progenitor channel to account for observable SGRBs. At the same time, the discovery of a new class of long-duration GRBs (LGRBs) that likely spawn from BNS mergers would serve to increase $\mathcal{R_{\rm BNS}}$ \citep{Rastinejad2022,Troja2022,Yang2022,Levan2024,Yang2024}, but the intrinsic rate of such events is highly uncertain.

We now revisit this comparison of rates in light of the updated BNS values in this paper, taking a value of $\mathcal{R_{\rm BNS}}\sim 100$~Gpc$^{-3}$~yr$^{-1}$ (\S~\ref{sec:GWrates}). Here, we define $\mathcal{R_{\rm SGRB}}$ as the true rate of SGRBs after corrections for beaming. In Figure~\ref{fig:rates_ratio}, we plot the ratio of rates, $\mathcal{R_{\rm SGRB}}/\mathcal{R_{\rm BNS}}$, as a function of jet opening angle $\theta_j$, in which we have assumed that the true rate is corrected from the observed one via the beaming factor, $f_b$ (i.e., $\mathcal{R_{\rm true}}=\mathcal{R_{\rm obs}}f_b^{-1}$) and that $f_b = 1-$cos($\theta_j$).\footnote{One caveat is that this assumption requires a top-hat jet orientation, as opposed to jet structure. If jet structure plays a role, this would only serve to enhance local rate estimates \citep{Howell2025}.} A horizontal line at unity represents the case in which $\mathcal{R_{\rm BNS}}=\mathcal{R_{\rm SGRB}}$. The parameter space {\it above} this line represents the scenario in which $\mathcal{R_{\rm BNS}}<\mathcal{R_{\rm SGRB}}$.

We first directly compare a variety of local SGRB rates (inferred within the LVK GW-detection volume of 200 Mpc; orange bands), in which the beaming correction becomes less significant for larger opening angles, thus resulting in a lower inferred true rate. Two of these local rates are inferred from the luminosity function of SGRBs \citep{Wanderman2015,Ghirlanda2016} assuming different minimum $\gamma$-ray luminosities, thus resulting in different rates. The third is from candidate off-axis SGRB afterglows within $<200$~Mpc (\citealt{Dichiara2020}; although see also: \citealt{Mandhai2018} who performed similar searches and did not find any viable candidates). We note that while several other local rates exist (i.e., \citealt{Nakar2006,GuettaStella2009,Coward2012,Zhang2018}), the ones selected here are representative of the range of rates in the literature (see also Fig. 1 in \citealt{Kunnumkai2026}). The width of the bands are driven by a combination of the uncertainties in the BNS rate and in the SGRB rates.

It is clear that most of the local rate measurements require wider jets of $\theta_j > 10^{\circ}$ to reconcile the BNS and SGRB rates. Figure~\ref{fig:rates_ratio} demonstrates that such wide jets are larger than the core of the jet from GW170817 (c.f., \citealt{MarguttiChornock2021}) and are only inferred for $\approx 30\%$ of the SGRB population with measured jets. Additionally, only four additional events have indications of wider jets, with lower limits of $\theta_j \gtrsim 10^{\circ}$ \citep{RoucoEscorial2023,Schroeder2025}. However, there can be intrinsic observational bias as wider jets can only be measured at later times post-SGRB, when the afterglows are already faint and often evade detection. We return to this point later in the section.

To investigate how the rates compare to those inferred outside of the local volume, and probe any redshift evolution in the rate, we also calculate $\mathcal{R_{\rm BNS}}/\mathcal{R_{\rm SGRB}}$ assuming rates inferred from cosmological SGRBs (red bands in Figure~\ref{fig:rates_ratio}). Here we adopt the range of rates published in \citet{RoucoEscorial2023} based on SGRB jet measurements of $\sim 360-1800$~Gpc$^{-3}$~yr$^{-1}$. These rates are a factor of $\sim 3.6-18$ {\it larger} than the central value of the latest BNS rate (100~Gpc$^{-3}$~yr$^{-1}$), depicted as the central red band in Figure~\ref{fig:rates_ratio}. For the most conservative and widest range, we also plot the ratio incorporating both the uncertainties on the SGRB and BNS rates (wider band in Figure~\ref{fig:rates_ratio}). For instance, using the 95\% upper limit on the BNS rate and the lower bound on the SGRB rate, to represent the lowest possible ratio, the SGRB rate is still $>1.3$ times larger than the BNS rate. We also investigate a range of other published SGRB rates and find that the large majority do not intersect with the latest median value of $\mathcal{R_{\rm BNS}}$ \rr{(e.g., the cosmological SGRB rates inferred by \citealt{2023A&A...680A..45S}, which although lower than \citealt{RoucoEscorial2023}, still prefer $\mathcal{R}_\mathrm{SGRB} > 100\,\mathrm{Gpc}^{-3}\,\mathrm{yr}^{-1}$)}. We uniformly find that not all SGRBs can be explained by BNS mergers.
\pnp{Our results are comparable to a recent study that found that low BNS rates $\mathcal{R}_\mathrm{BNS} =50 \,\mathrm{Gpc}^{-3}\,\mathrm{yr}^{-1}$ are in significant tension with the {\it Fermi}/GBM short GRB population -- regardless of jet structure~\citep{deSantis+26}. However, under the naive assumption of limited redshift evolution, we find an emergent tension even for $\mathcal{R}_\mathrm{BNS} =100 \,\mathrm{Gpc}^{-3}\,\mathrm{yr}^{-1}$.}

Indeed, this rates comparison implies that BNS mergers could only account for $\lesssim 10\%$ of SGRBs, and that another channel is required to produce a majority. While NSBH mergers can explain the energetics, spatial distributions, and host galaxy demographics of SGRBs (i.e., \citealt{Gompertz2020}), NSBH mergers similarly struggle to explain the rates, as the EM-bright mergers have an even lower rate than the overall BNS value (Section~\ref{sec:GWrates}). The total NSM rate, using conservative upper limits on the EM-bright NSBH merger rate, is a factor of $\sim 2-11$ smaller than the SGRB rate (assuming a central value of $147$~Gpc$^{-3}$~yr$^{-1}$). 

Another possibility is that there is contamination in what we classify as SGRBs from non-merger progenitors, thereby artificially raising the SGRB rate. Indeed, the classic delineation of $\sim2$~sec in $\gamma$-ray durations to separate short from long GRBs is imperfect and there are bound to be some SGRBs that genuinely originate from collapsars (i.e., the SGRB\,200826A with a duration of $1.1$~s but originated from a collapsar; \citealt{Ahumada2021}). However, the contamination would have to be $\gtrsim 90\%$; such a high rate of contamination is not supported by studies on $\gamma$-ray  \citep{Bromberg2013,Jespersen2020} or host galaxy properties \citep{Fong2022}. Moreover, the large majority of GRBs classified as short based on their $\gamma$-ray durations alone do not have associated supernovae \citep{Berger2014}.
Since kilonova emission is an indicator of an EM-bright (BNS or NSBH merger) progenitor, we can also use the fraction of kilonovae that have resulted from SGRBs as a proxy for the fraction of SGRBs that have resulted from NS mergers. This is complementary to kilonova-based rates from blind surveys \citep{Andreoni2020,Andreoni2021}, which broadly yield upper limits of $\lesssim 900$~Gpc$^{-3}$~yr$^{-1}$. Within a volume of $z<0.3$, the horizon to plausibly detect kilonovae similar to that of GW170817 with 8- to 10-meter class telescopes, there are 7~SGRBs with observations at the relevant timescales and to the relevant depths to detect kilonovae \citep{Rastinejad2021}. Of these 7, at least two have viable well-accepted kilonova candidates, indicating that at least $\approx 30\% \pm 16\%$ (using the Wilson score interval corresponding to 68\% confidence) of SGRBs originate from BNS mergers. We note that this number could be as high as 70\% if we include 3 more tentative kilonova candidates. Finally, the two likely kilonovae accompanying LGRBs at $z<0.1$ \citep{Rastinejad2022,Yang2022,Troja2022,Levan2024,Gillanders2025} indicate that some fraction of BNS or NSBH mergers are not be accounted for by local ``traditional'' SGRB rate estimates, which would only serve to further heighten tensions with GW-inferred rates. 

In light of this surprising implication, we re-visit the inferences on the cosmological SGRB rates inferred from SGRB jet opening angles. In \citet{RoucoEscorial2023}, a local rate of 10~Gpc$^{-3}$~yr$^{-1}$ was assumed as a baseline observed SGRB rate \citep{Nakar2006}. This observed rate was corrected to a range of true rates, set by (a) only SGRBs with measured opening angles (resulting in an upper estimate on the rate), and (b) SGRBs with measured opening angles and a mock population including wider jets (resulting in a lower estimate on the rate). If we instead employ a much lower local rate of 0.5~Gpc$^{-3}$~yr$^{-1}$ (representing the lowest local rate inferred for SGRBs; \citealt{Ghirlanda2016}) and re-calculate the rates assuming the two scenarios listed above, we find median rates of $\mathcal{R_{\rm SGRB}}\approx 18-89$~Gpc$^{-3}$~yr$^{-1}$. Here, the bounds are again set by the SGRBs with measured opening angles (upper bound) and a mock population including wider jets (lower bound). Thus, in order to fully reconcile the SGRB and BNS rates, this requires wider opening angles of $\gtrsim 10^{\circ}$ for the majority of events, a revision to local rates ($\lesssim 1$~Gpc$^{-3}$~yr$^{-1}$), substantial rate evolution resulting in larger cosmological rates, or a different progenitor for the large majority of SGRBs. Our conclusions are consistent with \citet{Kunnumkai2026}, who find that a local SGRB rate of $\sim10$~Gpc$^{-3}$~yr$^{-1}$ requires wide jet opening angles $\gtrsim 10^{\circ}$ in order to match the BNS rate, while narrow jet populations require lower SGRB local rates of $\lesssim 1$~Gpc$^{-3}$~yr$^{-1}$.

\section{R-Process}
\label{sec:rprocess}

In the aftermath of GW170817, many works estimated that essentially all of the MW \rprocess could be produced in binary neutron star mergers \citep[e.g.,][]{Rosswog2018}.
We revisit this calculation in light of the lower neutron star merger (NSM, which combines BNS and NSBH) rates from GWTC-4.

\subsection{Milky Way \rprocess Mass}

Previous estimates of the total MW \rprocess mass \citep[e.g.,][]{Shen2015,Hotokezaka2018,Rosswog2018} found a total \rprocess mass of ${\approx}19,000 M_\odot$.
Here we recalculate the total \rprocess mass for two main reasons:
(1) A large fraction of the MW's metals reside in its circumgalactic medium \citep{Tumlinson2017} which must be included when comparing to volumetric rates from transient surveys \citep[e.g.,][]{Maoz2025}. For example, velocity kicks drive some fraction of BNS metals to merge in the galaxy halo, which takes non-negligible time to return to the galaxy \citep{Nugent2025}.
(2) There are three \rprocess peaks, and in the Solar \rprocess pattern most of the mass (${\approx}90\%$) resides in the first peak \citep[e.g.,][]{Ji2019}. However, the Solar \rprocess patterns is calculated by subtracting off a model for the s-process \citep[e.g.,][]{Arlandini1999,Sneden2008}, and the first peak may be affected by details of the s-process model \citep[e.g.,][]{Bisterzo2014} or additional processes contributing to the first peak in the trans-iron region \citep{Schatz2022}. It is much more reliable to only consider elements above the second peak ($A \gtrsim 135$, $Z \geq 56$), where the \rprocess pattern is empirically universal \citep{Sneden2008}.
We thus recompute the total MW \rprocess mass including these two considerations.

We estimate the solar-metallicity-equivalent baryonic mass of the MW to be $11.9 \times 10^{10} \Msun$, about two times larger than previous calculations.
The main components of the baryonic mass are the stellar mass, the interstellar medium (ISM), and the circumgalactic medium (CGM). The total stellar mass is $5.4 \times 10^{10} \Msun$, and the ISM mass is ${\approx}$20\% of the stellar mass \citep{McMillan2017}. As with previous studies, we assume that all stars and ISM have the same \rprocess mass fraction as the Sun. The CGM mass in MW analogues is observationally found to be at least $6.5 \times 10^{10}\Msun$ \citep{Werk2014}, but at lower than Solar metallicity such that the amount of metals in the CGM is similar to or slightly more than those found in stars and the ISM \citep{Peeples2014,Tumlinson2017}.
We thus estimate that the total solar-metallicity-equivalent mass is $(1 + 0.2 + 1) = 2.2$ times larger than the stellar mass of the MW, or $11.9 \times 10^{10} \Msun$. This is consistent with estimates that half of all metals at $z=0$ reside in stars \citep{Peroux2020}.

\begin{figure}
    \centering
    \includegraphics[width=0.5\linewidth]{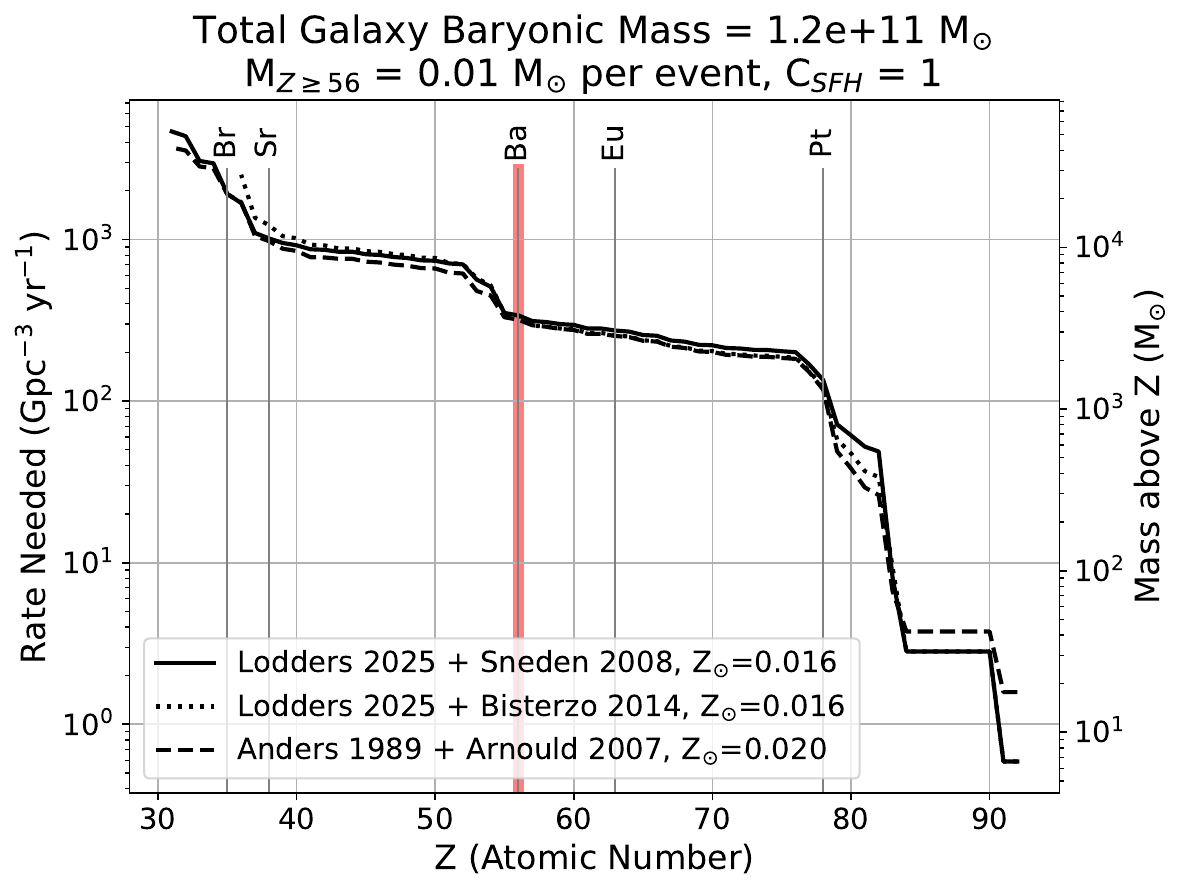}
    \caption{\rprocess event rate needed to reproduce the MW \rprocess mass. This includes material in both the stars and gas (interstellar and circumgalactic media). The right axis shows the results of our calculations for the for three different combinations of Solar isotopes/metallicities \citep{Anders1989,Lodders2025} and $r$/$s$ splits \citep{Arnould2007,Bisterzo2014}. The models agree very well for the heavier \rprocess elements with $Z \geq 56$. The left axis converts this total \rprocess mass into an implied volumetric rate assuming $0.01 M_\odot$ of heavy \rprocess material is synthesized in each event and $C_{\rm SFH} = 1$. 
    }
    \label{fig:rprocrateneeded}
\end{figure}

We redetermine the Solar \rprocess mass fractions empirically using a recent update to the Solar isotopic abundances by \citet{Lodders2025} and the $r$/$s$ contribution fractions in \citet{Sneden2008}, focusing only on isotopes with $Z \geq 56$ (or $A \gtrsim 135$) above the 2nd \rprocess peak. The mass fraction of these heavier \rprocess isotopes is $3.18 \times 10^{-8}$.
Combining the MW baryonic mass of $11.9 \times 10^{10} \Msun$ and the \rprocess mass fraction gives a total heavy \rprocess mass of $M_{r,Z\geq56} = 3784 \Msun$.
Our choice to use the empirical pattern for relatively massive \rprocess elements rather than a theoretical \rprocess calculation is due to substantial uncertainties in the nuclear data along the \rprocess path \citep{Mumpower2016,Schatz2022}. The universality of the \rprocess pattern between the 2nd and 3rd \rprocess peaks \citep{Sneden2008,Roederer2022} suggests that the empirical pattern is a sufficient approximation, and Figure~\ref{fig:rprocrateneeded} shows how robust the $r$/$s$ split is above $Z \geq 56$ \citep{Arnould2007,Bisterzo2014}. The primary uncertainty is thus dominated by the adopted Solar metallicity, where we adopt $Z_\odot=0.016$ but historically has varied by 15--25\%, from $Z_\odot = 0.013 - 0.020$ \citep{Anders1989,Asplund2009}.
We thus adopt $(3.18 \pm 0.7) \times 10^{-8}$ for the \rprocess mass fraction, which translates to $M_{r,Z\geq56} = 3784 \pm 833 \Msun$, which we round to $3800 \pm 800 \Msun$ going forward.

\subsection{Global \rprocess Production}

Following \citet{Rosswog2018} and \citet{Hotokezaka2018}, the total MW \rprocess production $M_r$ is given by:
\begin{equation}\label{eq:rprocproduced}
    M_r \sim 1120 M_\odot 
             \left(\frac{\mathcal{R}_{\text{BNS},z=0}}{100 \text{Gpc}^{-3} \text{yr}^{-1}}\right)
             \left(\frac{m_{ej}}{0.01 M_\odot}\right)
             \left(\frac{\tau_{\rm gal}}{13 \text{Gyr}}\right)
             \left(\frac{C_{\rm SFH}}{1}\right)
\end{equation}
where $\mathcal{R}_{\text{BNS},z=0}$ is the BNS rate at $z=0$ as measured by GW observatories,
$m_{ej}$ is the ejected \rprocess mass (here assumed to be the amount at $A \gtrsim 135$),
$\tau_{\rm gal}$ is the duration of star formation in the MW assuming a constant star formation rate,
and $C_{\rm SFH}$ is a factor accounting for a non-constant star formation rate.
The transformation from a volumetric to per-galaxy rate is done assuming $1.16 \times 10^{-2}$ Milky Way Equivalent Galaxies Mpc$^{-3}$\citep{Abadie2010}.
We can also rewrite this equation in terms of the BNS rate needed to produce all the \rprocess mass:
\begin{equation}\label{eq:rateneededforrproc}
    \mathcal{R}_{\text{BNS},z=0} \sim 338\, \text{Gpc}^{-3} \text{yr}^{-1}
             \left(\frac{M_{r,\text{needed}}}{3800 M_\odot}\right)
             \left(\frac{m_{ej}}{0.01 M_\odot}\right)^{-1}
             \left(\frac{\tau_{\rm gal}}{13 \text{Gyr}}\right)^{-1}
             \left(\frac{C_{\rm SFH}}{1}\right)^{-1}
\end{equation}
This equation is used to convert the \rprocess mass in Figure~\ref{fig:rprocrateneeded} to the volumetric rate needed.

The factor $C_{\rm SFH}$ in Eq.~\ref{eq:rprocproduced} and Eq.~\ref{eq:rateneededforrproc} accounts for any evolution in the the MW's NSM rate evolution. In the Solar Neighborhood, the MW's thin disk has a roughly constant star formation history going back to 10 Gyr ago \citep[e.g.,][]{Gallart2024}, which is the justification used for fixing $C_\mathrm{SFH} = 1$ in past versions of Eq.~\ref{eq:rprocproduced}. 
However, around 10 Gyr ago, there is a large burst of star formation forming over 80\% of the thick disk \citep[e.g.,][]{Xiang2025}. Even ignoring the bulge, ${\gtrsim}25$\% of the MW's stars form in this early burst since the thick disk is about 30\% of the thin disk mass \citep{McMillan2017}. If star formation peaks earlier than $z=0$, this implies a \textit{larger} total synthesized \rprocess mass in the MW over cosmic time compared to the $z=0$ NSM rate.

\begin{figure}
    \centering
    \includegraphics[width=0.5\linewidth]{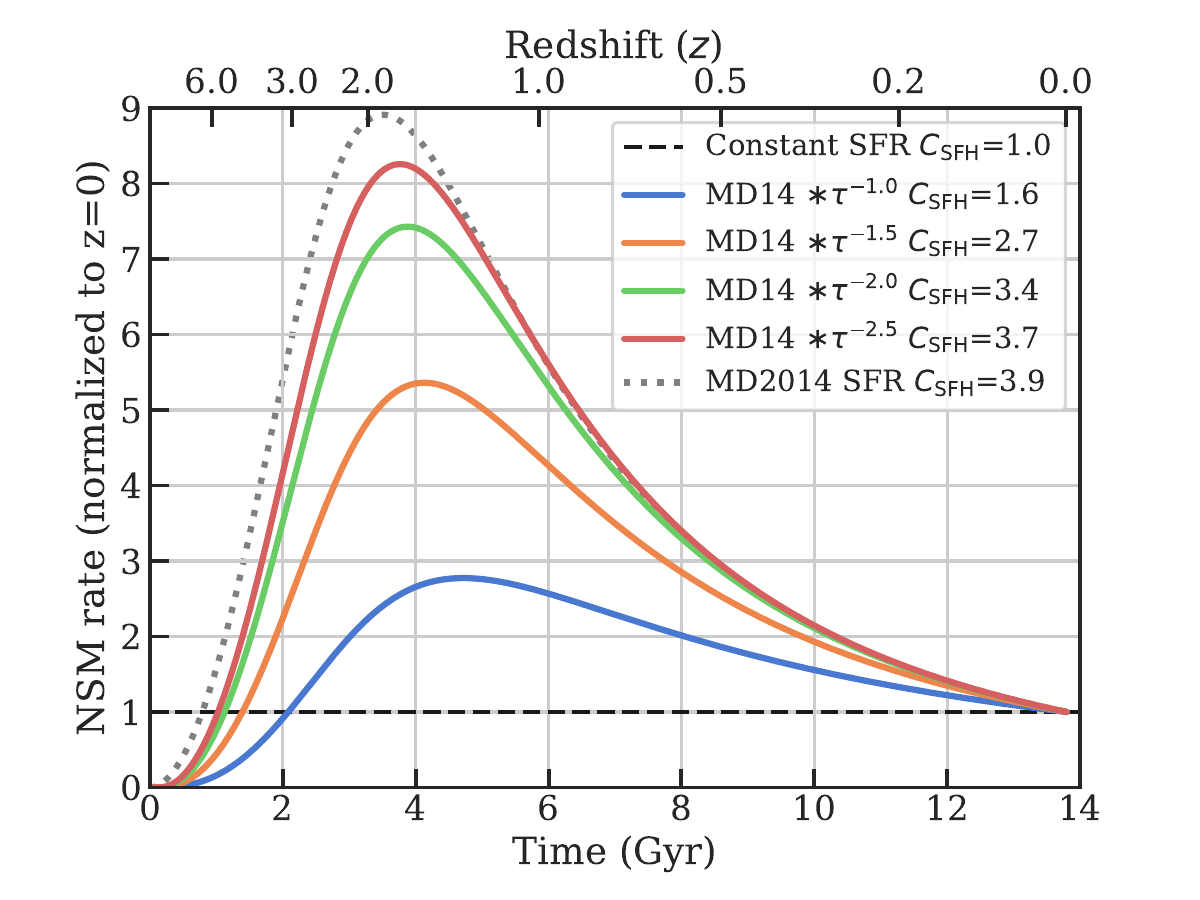}
    \caption{Neutron star merger rates over time for different delay time distributions, normalized to the $z=0$ rate. The dotted line shows the \citet{Madau2014} star formation rate (SFR), while the dashed line shows a constant SFR. The colored lines show the \citet{Madau2014} SFR convolved with different delay time distributions. More extended delay time distributions result in \textit{lower} integrated numbers of NSMs, because the $z=0$ rate is closer to the rate in the past. The legend shows $C_{\rm SFH}$ for different delay time distributions and star formation histories, where $C_{\rm SFH}$ is the enhancement in number of NSMs relative to $z=0$ due to the shape of the star formation history.}
    \label{fig:chemevolrates}
\end{figure}

In this work, we consider a range of $C_\mathrm{SFH}$ between 1 and 3, taking into account evolution of the MW's star formation rate and plausible variations in NSM delay time distribution. 
Figure~\ref{fig:chemevolrates} plots the time-evolving NSM rate normalized to $z=0$.
The dashed black horizontal line indicates a constant NSM rate ($C_\mathrm{SFH} = 1$) and the dotted grey shows the cosmic star formation history \citep[][MD14]{Madau2014}, which we use as a proxy for the MW's past star formation history. The colored lines show the time evolution of the NSM rate assuming the MD14 star formation history and different delay time distributions $\tau^{-\alpha}$ assuming a minimum delay time of 100 Myr.
The legend shows $C_{\rm SFH}$, the ratio in the integrated number of NSM produced over this time compared to a constant NSM rate. 
$C_{\rm SFH}$ is not very sensitive to different choices of the minimum delay time.

Our fiducial range $1 < C_{\rm SFH} < 3$ represents the uncertainty range between a constant star formation history (or long delay times $\alpha > -1$) and the MD14 SFR with steep delay times up to $\alpha \approx -2$.
$C_{\rm SFH} = 3$ roughly matches $\alpha=-1.83^{+0.35}_{-0.39}$, the slope inferred by \citet{Zevin2022} from modeling SGRB host galaxies. 
However, population synthesis simulations often predict shallower delay time distributions with $\alpha \approx -1$, and this is the distribution often adopted in chemical evolution models~\citep[e.g.][]{Kobayashi_2023}. This would correspond to $C_{\rm SFH} = 1.6$ assuming NSM progenitors follow the MD14 SFR.
Smaller values of $C_{\rm SFH}$ require the local NSM rate to be larger, \pnp{increasing the tension with the observed GW rate.}
Conversely, if nearly all NSMs occur through a very prompt channel, this increases $C_{\rm SFH}$ to 4, and the local NSM rate can be lower.

\rr{The major uncertainty other than $C_{\rm SFH}$ is the ejecta mass per merger $m_{\rm ej}$, where we adopted $m_{ej} = 0.01$.
First, recall that this is only the mass ejected of the heaviest \rprocess elements $Z \geq 56$ or $A \gtrsim 135$.
A simple empirical estimate for the ejecta mass of heavy \rprocess elements is to take the Solar \rprocess pattern's fraction for these heavy elements (${\lesssim}10\%$) and multiply by the total \rprocess ejecta mass inferred from kilonovae. For GW\,170817's kilonova, estimates of the lanthanide-rich component span from $\sim 0.01-0.1 M_\odot$ (e.g., \citealt{2017Natur.551...80K,Siegel+2019}). \citet{Rastinejad2025} find that, when uniformly modeled, GW\,170817's kilonova ejecta mass is comparable to the median of a sample of GRB kilonovae. Thus, we adopt $0.01\, M_\odot$ as an intermediate value.
This procedure may overestimate $m_{\rm ej}$, because the Solar fraction of heavy \rprocess elements is high compared to metal-poor \rprocess stars and initial estimates for GW170817 \citep{Ji2019}.}

\rr{Detailed simulations also suggest $m_{\rm ej} = 0.01\,M_\odot$ is reasonable but perhaps on the higher end. By restricting to $Z \geq 56$ and $A > 135$, we only include elements on the heavy side of the second \rprocess peak, which excludes essentially all ejecta with $Y_e \gtrsim 0.25$ \citep[e.g.,][]{Lippuner2015}. As a whole, merger simulations show that the total ejecta mass is positively correlated with electron fraction \citep[e.g.,][]{Just2015,Radice2018,Shibata2019,Fujibayashi2023,Just2023}, i.e. most of the ejected mass is from the first \rprocess peak as ejecta masses increase.
One major factor is the remnant NS's lifetime before collapsing to a black hole; a longer lived remnant produces more neutrinos, which both add energy to eject more mass but increase the electron fraction \citep[e.g.,][]{Lippuner2017}.
A second factor is the overall ratio of dynamical to disk ejecta, which is controlled by the merger mass ratio and eccentricity as well as the NS equation of state. Dynamical ejecta tend to have lower electron fractions, produce mostly the heavy \rprocess elements, and are almost always in the range $10^{-2}-10^{-3} M_\odot$ \citep[e.g.,][]{Radice2018}, while disk ejecta tend to dominate the total mass but the majority of the ejecta have higher $Y_e$ that do not make significant amounts of the heaviest \rprocess elements \citep[e.g.,][]{Fujibayashi2023,Just2023}.
Two recent end-to-end simulations predict dynamical ejecta of $0.5-1.0 \times 10^{-2} \,M_\odot$ and maximum 20\% extra contribution to low-$Y_e$ ejecta from the post-merger disk winds \citep{Fujibayashi2023,Just2023}.
}

In summary, we find that $3800 \pm 800\,M_\odot$ of \rprocess is needed, which is around twice as much as past calculations because a substantial amount of \rprocess is in the circumgalactic medium. On the other hand, the local NSM rate may be up to a factor of three lower if we account for the MW's declining star formation history. For a fiducial yield $m_{ej} = 0.01 M_\odot$ of heavy \rprocess per event and a range of $C_{\rm SFH}$ between 1 and 3, we require a NSM rate of $89$--$410\,\mathrm{Gpc}^{-3}\,\mathrm{yr}^{-1}$ to produce all of the \rprocess in the MW, where the lower end corresponds to steep delay time distributions and more steeply declining star formation rate. This estimate is still within the uncertainty range of the GW BNS rate, but trending towards the higher end.
\rr{The required NSM rate scales inversely with $m_\mathrm{ej}$, and our adopted value is conservatively high, leading to a conservatively low estimate for the implied NSM rate.}
\pnp{Our results are consistent with \citet{Kobayashi_2023}, \citet{Chen2025}, and \citet{saleem2026mergersfallshortnonmerger}, who found that prompt (i.e. short delay time) mergers are required to reconcile MW \rprocess observations with the local NSM rate probed by GWs. Such short delay time distributions are difficult to reconcile with population synthesis predictions.}

\section{Galactic Double Neutron Stars}
\label{sec:MW-DNS}

In addition to LVK detections, observations of Galactic DNS systems inform the NS mass distribution and DNS merger rate in a MW-like galaxy. 
The 10 known DNS systems that will merge within a Hubble time all have relatively low NS masses, with a mean total mass of $2.69\,M_\odot$ and a standard deviation of $0.12\,M_\odot$ \citep{LIGO2020}.
The total mass of GW170817 is consistent with this Galactic DNS mass distribution. 
However, the total mass of GW190425 exceeds the Galactic DNS distribution by at least 5 standard deviations \citep{LIGO2020}, suggesting the possibility of a distinct subpopulation that may be missing from the Galactic sample of not-yet-merged, radio bright systems, perhaps because such heavy binaries have faster merger times~\citep{Andrews:2019plw,Galaudage:2020zst} or avoid pulsar recycling~\citep{Vigna-Gomez:2021oqy}. 

If we assume that heavier binaries represent a distinct population that is missing from the Galactic observations, we would expect that the BNS merger rate inferred from Galactic systems would be consistent with the rate of GW170817-like mergers. The per-MW rate of low-mass, GW170817-like mergers is $5.3^{+17.6}_{-4.9}\,\mathrm{Myr}^{-1}$ (assuming a Milky Way Equivalent Galaxy density of $0.01\,\mathrm{Mpc}^{-3}$;~\citealt{Abadie2010}).
In order to estimate the BNS merger rate from Galactic DNS systems, one typically sums over the rate implied by each individual system, given by its inverse merger time, with a correction for pulsar beaming and radio survey selection effects, which consider pulsar luminosity and lifetimes~\citep{1991ApJ...380L..17P,Kim:2004kz}. 
An additional correction is sometimes included to account for the contribution from elliptical galaxies, assuming that the MW is representative of spiral galaxies~\citep{1991ApJ...380L..17P,2005CQGra..22S.935R,2008ApJ...675.1459K,2010ApJ...716..615O,Pol2019}; \rr{for example, \citet{2005CQGra..22S.935R} find that the contribution from elliptical galaxies should increase the local merger rate estimate by a factor of 2. This correction depends on the assumed delay time distribution (see Fig.~\ref{fig:chemevolrates}), with short delay times (i.e., if BNS mergers trace the star formation rate) leading to a smaller contribution from elliptical galaxies~\citep{1991ApJ...380L..17P}.}

Recent estimates of the BNS merger rate from Galactic DNS systems report $\mathcal{R}_{\rm MW} = 42^{+30}_{-14}\,\mathrm{Myr}^{-1}$ \citep{Pol2019}, assuming a beaming correction factor of $\sim4.6$ to correct for the solid angle of the pulsar's radio beam. 
Applying updated pulsar beaming correction factors, \citet{Grunthal2021} find a slightly lower merger rate of $\mathcal{R}_{\rm MW} = 32^{+19}_{-9}\mathrm{Myr}^{-1}$  at 90\% credibility.
This is on the higher end of the inferred GW170817-like rate ($5.3^{+17.6}_{-4.9}\mathrm{Myr}^{-1}$), but consistent within the 90\% credible intervals.
Including both observed BNS, the GW rate implies a MW equivalent rate of $11.0^{+19.2}_{-8.2}$ Myr$^{-1}$.
Hence, comparing the DNS rate to the total GW  BNS rate, under the assumption that the correction for DNS observational selection effects effectively accounts for the contribution of the missing ``heavy BNS" population in the MW, alleviates some of the tension.
If the GW-inferred BNS rate continues to be smaller than the MW DNS-inferred rate with improved statistical uncertainties, it may indicate that the MW is missing less DNS systems than typically assumed.
Compared to common assumptions, pulsars may be less beamed, radio surveys may be more complete, or the MW may host more BNS mergers than the typical Milky Way Equivalent Galaxy.

\section{Discussion}
\label{sec:discussion}
In the previous sections, we compared the local BNS merger rate as measured from GW observations to the merger rate inferred from the Galactic DNS population, the Galactic \rprocess mass, and the SGRB rate.
Table~\ref{tab:summary} summarizes these comparisons. 
We report the ratios between the BNS merger rate inferred by these other probes to the measured rate from GW observations, using the point estimate of $\mathcal{R}_\mathrm{BNS} = 100\,\mathrm{Gpc}^{-3}\,\mathrm{yr}^{-1}$ (within current 90\% uncertainties, the GW rate can be up to a factor of $\sim3$ higher or lower, but will most likely be lower post-O4 analysis; see also Fig.~\ref{fig:rate-comparisons}).
We then list the various physics assumptions that these rate ratios are sensitive to, and how realistic adjustments to these assumptions would alleviate or exacerbate the emerging tension with the low GW rate. 

\begin{deluxetable*}{llll}
\tablecaption{Summary of the comparisons between the inferred BNS merger rate and the SGRB, \rprocess and Galactic DNS rates. We highlight how varying our assumptions could either reduce the apparent tension in the rates $(\downarrow)$ or increase it $(\uparrow)$. \label{tab:summary}}
\tablehead{
\colhead{Change to Physics} & \colhead{$\mathcal{R}_{\rm SGRB}/\mathcal{R}_{\rm BNS}$} & \colhead{$\mathcal{R}_{\rm rproc\,needed}/\mathcal{R}_{\rm BNS}$} &
\colhead{$\mathcal{R}_{\rm DNS, MW}/\mathcal{R}_{\rm BNS}$}
}
\startdata
Default Values  & 
$3.6$--$18$ & 
0.9--4.1 & 2.3--5.1 \\ 
\cutinhead{Gamma Ray Burst Changes}
Wider Jet Angle & 
$\downarrow$ (Fig~\ref{fig:rates_ratio})& 
No change & No change \\
Lower local rate &
$\downarrow$ (Fig~\ref{fig:rates_ratio})& 
No change & No change \\
Add some Long GRBs &
$\uparrow$ &
No change & No change \\
\cutinhead{Total R-process Changes}
Lower MW R-Process Mass & 
No change & 
$\downarrow$ (Fig~\ref{fig:rprocrateneeded}) & No change \\
More R-Process Ejected per Merger & No change & $\downarrow$ & No change \\
Earlier MW Star Formation History & 
No change & 
$\downarrow$ (Fig~\ref{fig:chemevolrates}) & No change \\
\cutinhead{Galactic Double Neutron Star Changes}
Wider Pulsar Beaming & No change & No change & $\downarrow$ \\
MW BNS Rate Higher than Typical & No change & $\downarrow$ & $\downarrow$ \\
\cutinhead{NSM Population Changes}
Population of Very Low Mass NS (Fig.~\ref{fig:BNS-rate-v-mass}) & $\downarrow$ (factor of 2--3) & $\downarrow$ & No change (no such DNS observed)
\\
Only Low Spin NS (\S\ref{sec:GWrates-systematics}) & $\uparrow$ (factor of $\sim2$) & $\uparrow$ (factor of $\sim2$) & $\uparrow$ (factor of $\sim2$)
\\
Contribution from NSBH (\S\ref{sec:GWrates}) & $\downarrow$ (factor of $\lesssim1.5$) & $\downarrow$ & No change
\\
Steeper Delay Time Distribution & 
$\downarrow$ ($R_{z\sim0.5} > R_{z=0}$) &
$\downarrow$ (Fig~\ref{fig:chemevolrates}) & $\uparrow$ (larger correction for missing DNS) 
\enddata
\tablecomments{Adopting $\mathcal{R}_{\rm BNS}=100$ Gpc$^{-3}$ yr$^{-1}$}
\end{deluxetable*}

\subsection{Changes to gamma-ray burst assumptions}
The cosmological SGRB rate reported in \citet{RoucoEscorial2023} of $360$--$1800\,\mathrm{Gpc}^{-3}\,\mathrm{yr}^{-1}$ is a factor of $\sim3.6$--$18$ higher than the median BNS rate.
As Fig.~\ref{fig:rates_ratio} shows, assuming a wider SGRB jet opening angle corresponds to a lower astrophysical SGRB rate. According to most local SGRB measurements, matching the local SGRB rate to the BNS rate would require typical jet opening angles of {$\gtrsim 10^\circ$}, which is wider than most inferred jet opening angles, and seems unlikely even when accounting for observational bias that may favor the detection of afterglows from narrow jets. 

A lower local SGRB rate (inferred from SGRBs $< 200$ Mpc, before correcting for beaming), such as the \citet{Ghirlanda2016} value of $0.5\,\mathrm{Gpc}^{-3}\,\mathrm{yr}^{-1}$, would also lead to better agreement between the BNS and SGRB rates (Fig.~\ref{fig:rates_ratio}). However, this value is significantly lower than other estimates of the local SGRB rate. 

The tension between GRB and NSM rates worsens when we consider that a fraction of long GRBs may also originate from mergers. Based on the local {\it Swift} long GRB population, \citet{Troja2022} estimated that long GRBs derived from mergers could contribute up to $\sim 26\%$ (68\% confidence upper bound) of the local population. It has also been shown that the contamination goes both ways, and a fraction of SGRBs originate in collapsars rather than mergers (i.e., \citealt{Ahumada2021}), which would alleviate the rate tension. However, the simple rate comparisons implies that since only $\sim 5-25\%$ of SGRBs might be explained by BNS mergers (assuming a central value of $100$~Gpc$^{-3}$~yr$^{-1}$), and thus $\gtrsim 70\%$ could have non-merger origins. This is contradicted by the fraction of SGRBs with observed kilonovae counterparts and upper limits on SGRBs with supernova counterparts.

\subsection{Changes to \rprocess assumptions}
\label{sec:discussion-rprocess}
Our default assumptions in inferring the NSM rate from the MW \rprocess abundance lead to a rate that is a factor of $0.9$--$4.1$ greater than the median BNS rate. Within current uncertainties, this is not yet inconsistent with the BNS merger rate, but may require us to revisit our assumptions if the BNS rate is confidently measured below $89\,\mathrm{Gpc}^{-3}\,\mathrm{yr}^{-1}$ with future observations.

Our calculation used a MW \rprocess mass of {$3800\pm800\,M_\odot$}, focusing only on isotopes above the second \rprocess peak and taking into account the \rprocess in the circumgalactic medium.  
The main uncertainty in the \rprocess mass is the uncertain Solar metallicity.
A lower value of the MW \rprocess mass would require fewer NSM mergers and be compatible with a lower NSM rate (see Fig.~\ref{fig:rprocrateneeded}). 

To convert the MW \rprocess mass to a required NSM rate, we assumed each NSM ejects $0.01\,M_\odot$ of the heaviest \rprocess material.
A higher ejecta mass per merger would imply a lower required NSM rate (see Eq.~\ref{eq:rateneededforrproc}). However, this seems unlikely, because $m_{\rm ej} = 0.01\,M_\odot$ is on the high end of observational and theoretical estimates. 

Our calculation of the NSM rate required to explain the MW \rprocess mass is also sensitive to the NSM merger history, i.e., the MW star formation history convolved with NSM delay times via the factor $C_{\rm SFH}$ (see Fig.~\ref{fig:chemevolrates}). A flatter MW star formation history than we assumed here would imply fewer NSM mergers in the past, requiring more $z=0$ NSM mergers to explain the \rprocess mass. Conversely, low BNS merger rates suggest that $C_{\rm SFH}$ is relatively large, requiring a steeply declining MW star formation rate and/or prompt mergers.

The comparison between the MW \rprocess mass and the NSM rate also assumes that the MW is representative of $z=0$ galaxies, using a MW equivalent galaxy density to convert between MW rates and local rate densities. If the NSM rate in the MW is higher than typical, we would expect the inferred NSM rate from GWs to be lower than the MW rate. However, the opposite assumption is often made when converting the MW NSM rate to the local volumetric rate ~\citep{2008ApJ...675.1459K,Pol2019}.

\subsection{Changes to Galactic double neutron star assumptions}

The Galactic DNS systems imply a MW rate of $23$--$51\,{\rm Myr}^{-1}$~\citep{Grunthal2021}, or a volumetric rate of $230$--$510\,{\rm Gpc}^{-3}\,{\rm yr}^{-1}$ assuming a standard MW equivalent galaxy density.
Within uncertainties, this overlaps with our estimate of the total BNS merger rate, but suggests that we may soon need to revisit our assumptions.
The tension becomes more significant if we assert that only low-mass BNS mergers like GW170817 are representative of the merging DNS mass distribution, in which case the DNS rate should match the GW170817-like rate of $53^{+176}_{-49}\,\mathrm{Gpc}^{-3}\,\mathrm{yr}^{-1}$.

The inferred merger rate from the Galactic DNS systems would decrease if we assumed wider pulsar beaming, which would reduce the beaming correction factor. Similarly, assuming a higher survey completeness would reduce the correction for observational selection effects and correspond to a higher intrinsic DNS rate. 
As discussed above in \S\ref{sec:discussion-rprocess}, it may also be that the conversion from the MW rate to the volumetric rate through the assumed MW equivalent galaxy density is inaccurate.

\subsection{Changes to neutron star merger population assumptions}

The rate comparisons shown in Fig.~\ref{fig:rate-comparisons} and summarized in Tab.~\ref{tab:summary} use the inferred BNS merger rate. 
Nevertheless, a subpopulation of EM-bright NSBHs that include low-mass and/or spinning BHs could also contribute to \rprocess production and merger-origin GRBs. With the latest GW catalog, we only have upper limits on the EM-bright NSBH rate. Including a possible contribution from such NSBHs, the joint NSM rate is conservatively $55$--$350\,\mathrm{Gpc}^{-3}\,\mathrm{yr}^{-1}$, at most a factor of $\lesssim1.5$ higher than the BNS rate. 

In addition to a so far elusive population of EM-bright NSBH, there may be a population of very low-mass BNS (less massive than GW170817) that are missing in the GW sample because of the reduced sensitivity to low masses. As discussed in \S\ref{sec:GWrates-systematics} and shown in Fig.~\ref{fig:BNS-rate-v-mass}, taking into account the upper limit on the rate of BNS mergers with component masses $\lesssim1\,M_\odot$ could increase the BNS merger rate by a factor of 2--3. However, this is an extreme assumption, as such low mass NSs have never been observed. 

Our inferred NSM rates have also made the conservative, but perhaps unrealistic, assumption that BNS spins are uniformly distributed between dimensionless spin magnitudes of 0--0.4 and isotropically oriented. Current GW matched-filter searches only include templates with NS spins $<0.05$ and are therefore less sensitive to higher-spin BNS mergers. If NS spins in BNS mergers are all slowly spinning with dimensionless spins $<0.05$, consistent with the known spins of NSs in Galactic DNS systems, the inferred BNS merger rate would be a factor of $\sim2$ smaller than reported here, increasing the tension with the other rate estimates.

We also stress that current GW detectors can only probe the local, $z\sim0$ population of BNS mergers, whereas other probes may be sensitive to the past $z > 0$ or future $z < 0$ merger rate. 
If BNS mergers experience a steep delay time distribution, the merger rate is higher at $z > 0$. For example, the cosmological SGRB rate is sensitive to SGRBs at $z\sim0.5$, but the rates we use from \citet{RoucoEscorial2023} assume minimal evolution with redshift. There may be a factor of $\sim2$--3 increase in the merger rate at $z=0.5$ compared to $z=0$ from realistic delay time distributions, assuming BNS progenitors follow the star formation rate (see Fig.~\ref{fig:chemevolrates}). \pnp{Short delay times are indeed supported by recent analyses of the SGRB population~\citep{2023A&A...680A..45S, 2026arXiv260103861P,deSantis+26} and their host galaxies~\citep{Zevin2022}}.
Similarly, a steeper BNS delay time distribution would imply a greater production of \rprocess in the MW's past, requiring a lower $z=0$ BNS merger rate to match the MW's \rprocess mass (see the discussion about the $C_\mathrm{SFH}$ factor in \S\ref{sec:rprocess} and Fig.~\ref{fig:chemevolrates}). 
\pnp{Short delay times causing early \rprocess enrichment are also supported by the enrichment history of the MW, as probed by the relationship between europium and iron abundances~\citep[e.g.][]{Chen2025,saleem2026mergersfallshortnonmerger}.}
On the other hand, the Galactic DNS population is sensitive to the future merger rate, and a steep delay time distribution would imply a larger BNS merger rate at $z=0$ inferred from the DNS systems. In other words, the DNS systems with short delay times are missing from the observable DNS sample, and correcting for these missing systems would lead to a larger inferred BNS merger rate, increasing the tension with the measured BNS merger rate from GW observations.

\section{Conclusion}
\label{sec:conclusion}

Our main results and their interpretation are as follows:
\begin{itemize}

\item Observations through the first part of the LVK's fourth observing run imply a total BNS merger rate of $110^{+192}_{-82}\,\mathrm{Gpc}^{-3}\,\mathrm{yr}^{-1}$, with $53^{+176}_{-49}\,\mathrm{Gpc}^{-3}\mathrm{yr}^{-1}$ in GW170817-like $\sim(1.3,1.3)\,M_\odot$ BNSs, $30^{+99}_{-28}\,\mathrm{Gpc}^{-3}\mathrm{yr}^{-1}$ in heavy $\sim(1.65,1.65)\,M_\odot$ BNSs, and a conservative upper limit of $<31\,\mathrm{Gpc}^{-3}\,\mathrm{yr}^{-1}$ in heavier $\sim(2.1,2.1)\,M_\odot$ BNSs (see Fig.~\ref{fig:bns-rates}). If zero (one) new BNS events are announced following the full offline analysis of O4, we estimate that the total BNS rate would drop to approximately 16--$170\,\mathrm{Gpc}^{-3}\mathrm{yr}^{-1}$ (23--$253\,\mathrm{Gpc}^{-3}\mathrm{yr}^{-1}$) at 90\% credibility.

\item The comparison between the BNS and SGRB rate may point to a combination of wider jet angles ($\gtrsim 10^\circ$), lower local rates ($\sim 1\,\mathrm{Gpc}^{-3}\,\mathrm{yr}^{-1}$ rather than $\sim10\,\mathrm{Gpc}^{-3}\,\mathrm{yr}^{-1}$), alternative SGRB progenitors, and a negligible fraction of choked/ failed jets. It may also indicate a steep redshift evolution of the BNS merger rate between $z=0$ (as probed by current GW detectors) and $z\approx0.5$ (as probed with the cosmological SGRB rate), which could imply a combination of prompt BNS mergers and a preference for low-metallicity progenitors.

\item If future observations bring the NSM rate confidently below the rate required to produce the MW's \rprocess mass as inferred here ({89--$410\,\mathrm{Gpc}^{-3}\mathrm{yr}^{-1}$}), it may indicate a lower MW \rprocess mass than we assume, more than $0.01\,M_\odot$ of heavy \rprocess material ejected per merger, and/or an earlier SFH for the MW with a prompt BNS merger channel. It may also indicate a higher NSM rate in the MW than for a typical MW-like galaxy. 

\item The comparison between the merger rate inferred from Galactic DNS systems and the GW BNS rate may also suggest that the MW has a higher rate than typical galaxies, although this would contradict the correction that is commonly applied. It may also imply that the MW rate is overestimated due to, for example, wider pulsar beaming than assumed. Unlike for the GRB and \rprocess comparison, invoking a steep redshift evolution for BNS mergers would increase the tension with the BNS rate. 

\item The dearth of observed low-mass NSBH mergers implies a low rate of EM-bright NSBH mergers ($\lesssim50\%$ of the BNS rate), suggesting that they do not contribute significantly to the GRB rate and \rprocess production. Our assumptions about BNS masses and spins lead to a conservatively high BNS rate estimate. Realistic changes to the BNS spin distribution would imply a factor of 2 decrease in the BNS merger rate, while a hidden population of very low mass NS may lead to a factor of 2--3 increase in the rate. 

\end{itemize}




\begin{acknowledgments}
We are grateful to Teagan Clarke for thoughtful comments and suggestions on the manuscript. We also thank Sharan Banagiri, Tom Callister, Alexandra Guerrero, Daniel Holz, Paul Lasky, Samuele Ronchini, Javier Roulet, Nikhil Sarin and Michael Zevin for helpful discussions. 
MF acknowledges support from the Natural Sciences and Engineering Research Council of Canada (NSERC) under grant RGPIN-2023-05511, the Alfred P. Sloan Foundation, and the Ontario Early Researcher Award ER24-18-170.
A.P.J. acknowledges support from the National Science Foundation under grants AST-2307599 and AST-2510795, the U.S. Department of Energy Office of Nuclear Physics Award Number DE-SC0023128, and the Alfred P. Sloan Foundation.
W.F. gratefully acknowledges support by National Science Foundation under grant Nos. AST-2206494, AST-2308182, AST-2432037, and CAREER grant No. AST-2047919, the David and Lucile Packard Foundation and the Research Corporation for Science Advancement through Cottrell Scholar Award \#28284.
J. C. R. was supported by NASA through the NASA Hubble Fellowship grant \#HST-HF2-51587.001-A awarded by the Space Telescope Science Institute, which is operated by the Association of Universities for Research in Astronomy, Inc., for NASA, under contract NAS5-26555. H.-Y.C. is supported by the National Science Foundation under Grant PHY-2308752 and Department of Energy Grant DE-SC0025296.
This material is based upon work supported by NSF's LIGO Laboratory which is a major facility fully funded by the National Science Foundation.
\end{acknowledgments}

\bibliography{main}{}
\bibliographystyle{aasjournalv7}

\end{document}